\documentclass[aps,preprint,showpacs]{revtex4}
\usepackage{psfig}
\begin{document}        %
\draft
\title{Deformations of glassy polymers in very  low \\ temperature  regime  within cylindrical micropores}  %
\author{Guanghua Zhu} %\thanks{Corresponding author. E-mail address : ).}}
\affiliation{Chern S.-S. Institute of Mathematics, Nankai
University, Road Weijin, Tianjin 300071, China China}
%\begin{document}
%
\begin{abstract}
The  deformation kinetics for glassy polymers confined in
microscopic domain at very low temperature regime was investigated
using a transition-rate-state dependent model considering the
shear thinning behavior which means, once material being subjected
to high shear rates, the viscosity diminishes with increasing
shear rate. The preliminary results show that there might be
nearly frictionless fields for rate of deformation due to the
almost vanishing shear stress in micropores at very low
temperature regime subjected to some surface conditions : The
relatively larger roughness (compared to the macroscopic domain)
inside micropores and the slip. As the pore size decreases, the
surface-to-volume ratio increases and therefore, surface roughness
will greatly affect the deformation kinetics in micropores. By
using the boundary perturbation method, we obtained a class of
temperature and activation energy dependent fields for the
deformation kinetics at low temperature regime with the presumed
small wavy roughness distributed along the walls of an cylindrical
micropore. The critical  deformation kinetics of the glassy matter
is dependent upon the temperature, activation energy, activation
volume, orientation dependent and is proportional to the
(referenced) shear rate, the slip length, the amplitude and the
orientation of the wavy-roughness.
%\newline
%
%\noindent {\it Keywords} : Thermomechanical processing; Amorphous
%Materials; Strain rate sensitivity;  Activation volume; Activation
%energy; Micromechanical Modeling
%
%
\end{abstract}
\pacs{83.60.St, 83.60.Rs,83.50.Lh}
%,67.80.-s, 67.40.Hf, 68.35.Ct, 47.37.+q, 47.61.-k, }
%----------------------------------------------------------------------
%
%\doublerulesep=7mm        %\parskip=12 pt
%\baselineskip=7mm \oddsidemargin-1mm
\maketitle
\bibliographystyle{plain}
%{\bf
\section{Introduction}
Glasses are amorphous materials of polymeric, metallic, inorganic
or organic type. The plastic deformation of amorphous materials
and glasses at low temperatures and high strain rates is known to
be inhomogeneous and rate-dependent. In fact, the mechanical
behavior of amorphous materials such as polymers [1-7] and bulk
metallic glasses [8-10] continues to present great theoretical
challenges. While dislocations have long been recognized as
playing a central role in plasticity of crystalline systems, no
counterpart is easily identifiable in disordered matter. In
addition, yield and deformation kinetics [11-13] occur very far
from equilibrium, where the state of the system may have a complex
history dependence.
\newline In recent years, considerable effort was geared towards
understanding how glasses respond to shear [14]. Phenomena such as
shear thinning and 'rejuvenation' are common when shear
deformation (rate) is imposed. At low temperatures they behave in
a brittle elastic manner; at high temperatures, much above the
glass transition the behavior is more (rubbery) (viscoelastic).
There is a huge drop in modulus when the temperature is increased
above the glass transition temperature, indicating a shift in
behavior from (glassy) to (rubbery). Because of these peculiar
mechanical properties of polymeric materials, the linear theory of
viscoelasticity is unable to model closely the observed response
and thus there is a need for a non-linear theory of
viscoelasticity. \newline Unlike crystals, glasses also age,
meaning that their state depends on their history. When a glass
falls out of equilibrium, it evolves over very long time scales.
Motivated by the above issues and the interesting characteristics
of deformation kinetics at very low temperature  we shall study
the deformation kinetics in microscopic domain at low enough
temperature which is an interesting topic for applications in
micro- and nanodomain [15-16] or the validation in using quantum
mechanic formulations [9,17] where the nonlinear constitutive
relations should be adopted.
\newline Meanwhile most of the classical solutions of contact
problems, starting from the Hertzian case, rely on the assumption
of nominally smooth geometries, which is reasonable at large
enough scales. However, real surfaces are rough [18-19] at the
micro- or even at the meso-scale, and the effect of multi-point
contact is important for a series of phenomena involving friction
and wear. The role of surface roughness has been extensively
investigated, and opposite conclusions have been reached so far.
For instance, friction can increase when two opposing surfaces are
made smoother (this is the case of cold welding of highly polished
metals). On the other hand, friction increases with roughness when
interlocking effects among the asperities come into play. This
apparent contradiction is due to the effects of length scales,
which appear to be of crucial importance in this phenomenon
[15,20-21].
\newline From the mechanical point of view, a contact problem
involves the determination of the traction distributions
transmitted from one surface to the other, in general involving
normal pressures and, if friction is present, shear tractions,
according to an appropriate set of equalities and inequalities
governing the physics of the contact [22]. When there is friction
at the contact interface, Coulomb friction behaviour is usually
introduced to give the conditions necessary to determine the shear
traction distribution. Any point in the contact area must be
either in 'stick', or 'slip' condition, and the tangential
tractions must behave accordingly.\newline
%---------------------------
%Recent results show that
%plastic deformation in solid helium creates defects and pressure
%gradients which are not easily eliminated by thermal
%annealing$^{\cite{Day:Melt}}$. Meanwhile these recent experiments
%have shown that solid helium does not flow in response to pressure
%gradients at low temperatures. However, close to the melting
%temperature  mass flow was indeed observed, but it decreases
%rapidly with temperature$^{\cite{Day:Melt}}$. For solid helium in
%the pores of Vycor the flow appears to be thermally activated and
%disappears below about half the melting temperature.
%-----------
%As proposed by Andreev [7] and Rittner and Reppy [4] the observed supersolidity
%might be in a glassy solid (helium) state [8]. Both idea could be inspired from the
%annealing effect of supersolidity reported before [9] and the recent experimental report
%of the existence of glassy supersolid helium.
In this paper we shall consider the deformation kinetics of glassy
polymer at very low temperature in micropores which have radius-
or transverse-corrugations along the cross-section. The glassy
polymer will be treated as a shear-shinning material. To consider
the transport of this kind of glass (shear-thinning) polymer in
microdomain, we adopt the verified model initiated by Cagle and
Eyring [8] which was used to study the annealing of glass. To
obtain the law of annealing of glass for explaining the too rapid
annealing at the earliest time, because the relaxation at the
beginning was steeper than could be explained by the bimolecular
law, Cagle and Eyring [8] tried a hyperbolic sine law between the
shear (strain) rate : $\dot{\xi}$ and (large) shear stress :
$\tau$ and obtained the close agreement with experimental data.
This model has sound physical foundation from the thermal
activation process (Eyring [9] already considered a kind of
(quantum) tunneling which relates to the matter rearranging by
surmounting a potential energy barrier; cf., Fig. 1).
%and thus it could resolve the concern raised by Anderson [12] for the thermal noises to the superflow of
%vortex liquid (i.e., the supersolid helium).
\noindent With this model we can associate the (glassy) polymer
with the momentum transfer between neighboring atomic clusters on
the microscopic scale and reveals the atomic interaction in the
relaxation of flow with (viscous) dissipation. \newline The
outline of this short paper is as follows. Section 2 describes the
general mathematical and physical formulations of the framework.
In this Section, explicit derivations  for the glassy deformation
kinetics are introduced based on a  microscopic model proposed by
Eyring [9]. The boundary perturbation technique [18,23] will be
implemented, too. In the third Section, we consider the very-low
temperature limit of  our derived solutions which are highly
temperature as well as activation energy dependent at rather low
temperature regime. Relevant results and discussion are given
therein.
%------------------------------
\section{Mathematical and physical formulations}
The beginnings of theoretical molecular mechanisms of deformation
in amorphous polymers and glass are as old as the subject of
atomic mechanisms of deformation and yield in metals. The first
specific molecular mechanism of deformation for polymers and glass
was published by Eyring [9] and later, Taylor [24] published the
model of an edge dislocation to account for the plastic
deformation in metals. However, whilst the theory of dislocations
and crystal defects has become a major stream in the science of
solid state, the corresponding effort applied to this problem in
amorphous polymers must be considered rather small by comparison.
\newline
%--------
The molecular theory of deformation kinetics came from a different
stream of science than that of structure and motion of crystal
defects (in particular dislocations). Its roots stretch to the
developmental stages of theories of chemical reactions and
thermodynamic description of their temperature dependence,
culminating in the key formulation by Arrhenius of the equation
for reaction rates. By the beginning of this century the concept
of activation entropy was included in the model, and it was
considered that molecules go both in the forward direction
(product state) and in the backward direction (reactant state).
\newline The development of statistical mechanics, and later quantum
mechanics, led to the concept of the potential energy surface.
This was a very important step in our modem understanding of
atomic models of deformation. Eyring's contribution to this
subject was the formal development of the transition state theory
which provided the basis for deformation kinetics, as well as all
other thermally activated processes, such as crystallisation,
diffusion, polymerisation. etc. \newline
%---------
The motion of atoms is represented in the configuration space; on
the potential surface the stable molecules are in the valleys,
which are connected by a pass that leads through the saddle point.
An atom at the saddle point is in the transition (activated)
state. Under the action of an applied stress the forward velocity
of a (plastic) flow unit is the net number of times it moves
forward, multiplied by the distance it jumps. Eyring proposed a
specific molecular model of the amorphous structure and a
mechanism of deformation kinetics. With reference to this idea,
this mechanism results in a (shear) strain rate given by
\begin{equation}  %(Eq. (16))
 \dot{\xi}=2\frac{V_h}{V_m}\frac{k_B T}{h}\exp (\frac{-\Delta
E}{k_B T}) \sinh(\frac{V_h \tau}{2 k_B T})
\end{equation}
where $$V_h=\lambda_2\lambda_3\lambda, \hspace*{24mm}
V_m=\lambda_2\lambda_3\lambda_1,$$ $\lambda_1$ is the
perpendicular distance between two neighboring layers of molecules
sliding past each other, $\lambda$ is the average distance between
equilibrium positions in the direction of motion, $\lambda_2$ is
the distance between neighboring molecules in this same direction
(which may or may not equal $\lambda$), $\lambda_3$ is the
molecule to molecule distance in the plane normal to the direction
of motion, and $\tau$ is the local applied stress, $\Delta E$ is
the activation energy, $h$ is the Planck constant, $k_B$ is the
Boltzmann constant, $T$ is the temperature, $V_h$ is the
activation volume for the molecular event [9]. The deformation
kinetics of the polymer chain is envisaged as the propagation of
kinks in the molecules into available holes. In order for the
motion of the kink to result in a plastic flow, it must be raised
(energised) into the activated state and pass over the saddle
point. This was the earliest molecular theory of yield behaviour
in amorphous polymers, and Eyring presented a theoretical
framework which formed the basis of many subsequent
considerations.
\newline
%Equation (1) is frequently modified under the condition
%$V_h f/(k_B T) \ge 1$ and the hyperbolic sine function is replaced
%by an exponential:
%$$  \dot{\xi}= A \exp (\frac{-\delta G +V_h f}{k_B T})  $$
%where $A$ encompasses all of the pre-exponential factors, $V_h$ is
%now referred to as the 'activation volume', and $f$ as the applied
%shear stress. However, formally Eyring did not distinguish between
%viscous and plastic deformation, and Eqn. (1) has been applied
%equally to both modes of deformation.
Solving Eqn. (1) for the force or $\tau$, one obtains:
\begin{equation}
  \tau=\frac{2 k_B T}{V_h} \sinh^{-1} (\frac{\dot{\xi}}{B}),
\end{equation}
which in the limit of small $(\dot{\xi}/B)$ reduces to Newton's
law for viscous deformation kinetics.
%The 'plastic' character only manifests itself when the magnitude of the applied stress times the
%activation volume becomes comparable or greater in magnitude than
%the thermal vibrational energy.
\newline
\noindent We  consider a steady deformation kinetics of the glassy
polymer in a wavy-rough microtube of $r_o$ (in mean-averaged outer
radius) with the outer wall being a fixed
wavy-rough surface : $r=r_o+\epsilon \sin(k \theta)$ %and $r_i$ (in
%mean-averaged inner radius) with the inner wall being a fixed
%wavy-rough surface : $r=r_i+\epsilon \sin(k \theta+\beta)$,
where $\epsilon$ is the amplitude of the (wavy) roughness, and the
wave number : $k=2\pi /L $ ($L$ is the wave length). The schematic
is illustrated in Fig. 2. Firstly, this material can be expressed
as [9,18]
%\begin{displaymath}
 $\dot{\xi}=\dot{\xi}_0  \sinh(\tau/\tau_0)$,
%\end{displaymath}
where $\dot{\xi}$ is the shear rate, $\tau$ is the shear stress,
and
\begin{equation}
\dot{\xi}_0 \equiv B= \frac{2 k_B T}{h}\frac{V_h}{V_m} \exp
(\frac{-\Delta E}{k_B T}),
\end{equation}
 is a function of temperature with the
dimension of the shear rate,
\begin{equation}
\tau_0 =\frac{2 k_B T}{V_h}
\end{equation}
 is the referenced (shear) stress, (for
small shear stress $\tau \ll \tau_0$, the linear dashpot
constitutive relation  is recovered and $\tau_0/\dot{\xi}_0$
represents the viscosity of the material). In fact, the force
balance gives the shear stress at a radius $r$ as $\tau=-(r
\,dp/dz)/2$. $dp/dz$ is the pressure gradient along the tube-axis
or $z$-axis direction.\newline Introducing the forcing parameter
%\begin{displaymath}
$\Pi = -(r_o/2\tau_0) dp/dz$
%\end{displaymath}
then we have
%\begin{displaymath}
 $\dot{\xi}= \dot{\xi}_0  \sinh ({\Pi r}/{r_o})$.
%\end{displaymath}
As the (shear) strain rate is
\begin{equation}
\dot{\xi}= \frac{du}{dr}
\end{equation}
($u$ is the rate of deformation (or velocity)  in the longitudinal
($z$-)direction of the microtube), after integration, we obtain
\begin{equation}
 u=u_s +\frac{\dot{\xi}_0 r_o}{\Pi} [\cosh \Pi - \cosh (\frac{\Pi r}{r_o})],
\end{equation}
here, $u_s$ is the rate of deformation or velocity over the
surfaces of the microtube, which is determined by the boundary
condition. We noticed that
%the most well-known slip-velocity boundary condition is the linear
%Navier relation [13], $\Delta u=L_s^0 \dot{\xi}$, in which
 a general boundary condition for material deformation kinetics over a solid
surface was proposed (cf., e.g.,  [18]) as
\begin{equation}
 \delta u=L_s^0 \dot{\xi}
 (1-\frac{\dot{\xi}}{\dot{\xi}_c})^{-1/2},
\end{equation}
where $\delta u$ is the rate of deformation (or velocity) jump
over the solid surface, $L_s^0$ is a constant slip length and
$\dot{\xi}_c$ is the critical shear rate at which the slip length
diverges. The value of $\dot{\xi}_c$ is a function of the
corrugation of interfacial energy. We remind the readers that this
expression is based on the assumption of the shear rate over the
solid surface being much smaller than the critical shear rate of
$\dot{\xi}_c$. $\dot{\xi}_c$ represents the maximum shear rate the
material can sustain beyond which there is no additional momentum
transfer between the wall and material-flow molecules. How generic
this behavior is and whether there exists a comparable scaling for
polymeric or amorphous materials remain open questions.\newline At
small pressure gradient, the shear-thinning matter behaves like a
Newtonian flow, while at high pressure gradient, the
shear-thinning matter flows in a plug-flow type. Such a behavior
is due to the shear thinning of the material, i.e., the higher the
shear rate is, the smaller is the (plastic) flow resistance [7].
On the microscale, this shear-thinning matter can bridge the
Newtonian deformation kinetics to that of the pluglike type and
offers us a mechanistic model to study the deformation kinetics in
micro- and even nanodomain
 using the technique of continuum
mechanics.
\newline % %$u=K_n du/dn \mid_r$;
With the boundary condition from (cf., e.g.,  [18]), we shall
derive the rate of deformation (or velocity) field or deformation
kinetics along the wavy-rough microtube below using the boundary
perturbation technique (cf. [23]) and dimensionless analysis. We
firstly select the hydrodynamical diameter $L_r$ to be the
characteristic length scale and set
\begin{equation}
r'=r/L_r, \hspace*{6mm} R_o=r_o/L_r, %\hspace*{6mm} R_i=r_i/L_r,
\hspace*{6mm} \epsilon'=\epsilon/L_r.
\end{equation}
After this, for simplicity, we drop all the primes. It means, now,
$r$, $R_o$, $R_i$, and $\epsilon$ become dimensionless. The wall
is prescribed as $r=R_o+\epsilon \sin(k\theta)$, %$r=R_i+\epsilon
%\sin(k \theta+\beta)$
and the presumed fully-developed plastic flow is along the
$z$-direction (microtube-axis direction). Along the confined
(wavy) boundaries, we have the strain rate
\begin{equation}
 \dot{\xi}=(\frac{d u}{d n})|_{{\mbox{\small on surface}}},
\end{equation}
where, $n$ means the  normal. Let the rate of deformation $u$ be
expanded in $\epsilon$ :
\begin{equation}
 u= u_0 +\epsilon u_1 + \epsilon^2 u_2 + \cdots,
\end{equation}
and on the boundary, we expand $u(r_0+\epsilon dr,
\theta(=\theta_0))$ into
\begin{displaymath}
u(r,\theta) |_{(r_0+\epsilon dr ,\theta_0)}
=u(r_0,\theta)+\epsilon [dr \,u_r (r_0,\theta)]+ \epsilon^2
[\frac{dr^2}{2} u_{rr}(r_0,\theta)]+\cdots=
\end{displaymath}
\begin{equation}
 \hspace*{12mm} \{u_{slip} +\frac{\dot{\xi} R_o}{\Pi} \cosh
 (\frac{\Pi \bar{r}}{R_o})|_r^{R_o+\epsilon \sin(k\theta)},
 \hspace*{6mm} r_0 \equiv R_o;
\end{equation}
where the subscript means the partial differentiation (say, $u_r
\equiv
\partial u/\partial r$) and
\begin{equation}
 u_{slip}|_{{\mbox{\small on surface}}}=L_s^0 \dot{\xi} [(1-\frac{\dot{\xi}}{\dot{\xi}_c})^{-1/2}]
 |_{{\mbox{\small on surface}}},
\end{equation}
\begin{equation}
 u_{{slip}_0}= L_s^0 \dot{\xi}_0 [\sinh\Pi(1-\frac{\dot{\xi}_0 \sinh\Pi}{
 \dot{\xi}_c})^{-1/2}].
\end{equation}
Now, on the outer wall (cf., e.g., [23]), the (shear) strain rate %\[
\begin{displaymath}
 \dot{\xi}=\frac{du}{dn}=\nabla u \cdot \frac{\nabla (r-R_o-\epsilon
\sin(k\theta))}{| \nabla (r-R_o-\epsilon \sin(k\theta))
|}=[1+\epsilon^2 \frac{k^2}{r^2}  \cos^2 (k\theta)]^{-\frac{1}{2}}
[u_r |_{(R_o+\epsilon dr,\theta)} -
%\]
\end{displaymath}
\begin{displaymath}  %\[
 \hspace*{12mm} \epsilon \frac{k}{r^2}
\cos(k\theta) u_{\theta} |_{(R_o+\epsilon dr,\theta)}
]=u_{0_r}|_{R_o} +\epsilon [u_{1_r}|_{R_o} +u_{0_{rr}}|_{R_o}
\sin(k\theta)-
\end{displaymath}
\begin{displaymath}
  \hspace*{12mm}  \frac{k}{r^2} u_{0_{\theta}}|_{R_o} \cos(k\theta)]+\epsilon^2 [-\frac{1}{2} \frac{k^2}{r^2} \cos^2
(k\theta) u_{0_r}|_{R_o} + u_{2_r}|_{R_o} + u_{1_{rr}}|_{R_o} \sin(k\theta)+ %\end{displaymath} %\]
\end{displaymath}
\begin{equation}
   \hspace*{12mm} \frac{1}{2} u_{0_{rrr}}|_{R_o} \sin^2 (k\theta) -\frac{k}{r^2}
\cos(k\theta) (u_{1_{\theta}}|_{R_o} + u_{0_{\theta r}}|_{R_o}
\sin(k\theta) )] + O(\epsilon^3 ) .
\end{equation}
%We consider $1 \gg K_n \gg \epsilon$ case,
Considering $L_s^0 \sim R_o \gg \epsilon$ case, we presume
$\sinh\Pi \ll \dot{\xi}_c/\dot{\xi_0}$ so that we can
approximately replace
%\begin{displaymath}
$[1-(\dot{\xi}_0 \sinh\Pi)/\dot{\xi}_c]^{-1/2}$
%\end{displaymath}
by
%\begin{equation}
$[1+\dot{\xi}_0 \sinh\Pi/(2 \dot{\xi}_c)]$.
%\end{equation}%--------------- Rough
With equations (6),(7),(9), (10), (11) and (14), using the
definition of the (shear) strain rate $\dot{\xi}$, we can derive
the rate of deformation (or velocity) field up to the second
order. The key point is to firstly obtain the slip rate of
deformation (or velocity) along the wavy boundaries or surfaces.
\newline After lengthy mathematical manipulations and using
%\begin{displaymath}
 $(1-{\dot{\xi}}/{\dot{\xi}_c})^{-1/2}\approx 1+{\dot{\xi}}/({2
 \dot{\xi}_c})$,
%\end{displaymath}
\begin{equation}
 u_0=-\frac{\dot{\xi}_0 R_o}{\Pi} [\cosh
 (\frac{\Pi r}{R_o})-\cosh \Pi]+u_{{slip}_0},
\end{equation}
\begin{equation}
  u_1=
 \dot{\xi}_0 \sin (k\theta) \sinh \Pi +u_{{slip}_1},
\end{equation}
we have %
%\vspace{13mm}
\begin{displaymath}
 u_{slip}=L_s^0 \{[-u_{0_r}(1-\frac{u_{0_r}}{2
 \dot{\xi}_c})]|_{r=R_o}+\epsilon
 [-u_f(1-\frac{u_{0_r}}{\dot{\xi}_c})]|_{r=R_o}+\epsilon^2
 [\frac{u_f^2}{2\dot{\xi}_c}-u_{sc}
 (1-\frac{u_{0_r}}{\dot{\xi}_c})]|_{r=R_o}\}=
\end{displaymath}
\begin{equation}
 \hspace*{36mm} u_{slip_0} +\epsilon \,u_{slip_1} + \epsilon^2 u_{slip_2}
 +O(\epsilon^3)
\end{equation}
where
\begin{equation}
 u_{0_r}= -\dot{\xi}_0 \sinh(\frac{\Pi}{R_o}r),
\end{equation}
\begin{equation}
 u_{0_{rr}}=-\dot{\xi}_0 \frac{\Pi}{R_o} \cosh(\frac{\Pi}{R_o}r),
\end{equation}
\begin{equation}
 u_{0_{rrr}}=-\dot{\xi}_0
\frac{\Pi^2}{R_o^2}\sinh(\frac{\Pi}{R_o}r),
\end{equation}
\begin{equation}
 u_f =u_{1_r} + u_{0_{rr}} \sin (k\theta)-\frac{k}{r^2} \cos (k\theta)
 u_{0_{\theta}}=-\frac{\Pi}{R_o}\dot{\xi}_0 \cosh(\frac{\Pi}{R_o}r) \, \sin (k\theta),
\end{equation}
and
\begin{equation}
 u_{sc} =-\frac{k^2}{2 r^2}\cos^2 (k \theta) u_{0_r} +\frac{1}{2}
 u_{0_{rrr}} \sin^2 (k\theta)=\frac{1}{2}\dot{\xi}_0 [\frac{k^2}{2 r^2}\cos^2 (k \theta)
 -\frac{\Pi^2}{R_o^2}\sin^2 (k\theta)]\sinh(\frac{\Pi}{R_o}r).
\end{equation}
Thus, at $r=R_o$, up to the second order,
\begin{displaymath}
 u_{slip}\equiv u_s=L_s^0 \dot{\xi}_0  \sinh\Pi(1+\frac{K_0}{2})+\epsilon \dot{\xi}_0 \sin(k\theta)
 [\sinh \Pi+ \frac{\Pi}{R_o}L_s^0\cosh\Pi \, (1+K_0)]+\epsilon^2 L_s^0\frac{\dot{\xi}_0 }{2}\{
 [
\end{displaymath}
\begin{equation}
 \frac{\Pi \cosh \Pi}{R_o L_s^0}  \sin^2 (k\theta)-\frac{k^2}{R_o^2} \cos^2 (k\theta)+
 \frac{\Pi^2}{R_o^2} \sin^2 (k\theta)]\sinh\Pi (1+K_0) +
 \frac{\Pi^2}{R_o^2} \frac{\dot{\xi}_0}{\dot{\xi}_c}
  \cosh^2 \Pi \,\sin^2 (k\theta) \},
% \epsilon^2 [-\frac{k^2}{2 a^2}\cos^2 (k \theta) \dot{\xi} \sinh \xi+
\end{equation}
where
\begin{equation}
K_0=1+({\dot{\xi}_0
 \sinh\Pi})/{\dot{\xi}_c}
\end{equation}
From the rate of deformation (or velocity) fields (up to the
second order), we can integrate them with respect to the
cross-section to get the volume (plastic) flow rate ($Q$, also up
to the second order here).
\begin{equation}
  Q=\int_0^{\theta_p} \int^{R_o+\epsilon \sin(k\theta)}
 u(r,\theta) r
 dr d\theta =Q_{smooth} +\epsilon\,Q_{p_0}+\epsilon^2\,Q_{p_2}.
% \newline
\end{equation}
\section{Results and discussion}
We firstly check the roughness effect upon the shearing
characteristics because there are no available experimental data
and numerical simulations for the same geometric configuration
(microscopic tubes with wavy corrugations in transverse
direction). With a series of forcings (due to imposed pressure
gradients) : $\Pi\equiv R_o (-dp/dz)/(2\tau_0)$, we can determine
the enhanced shear rates ($d\xi/dt$) due to forcings. From
equation (5), we have (up to the first order)
\begin{equation}
 \frac{d\xi}{dt}=\frac{d\xi_0}{dt} [ \sinh \Pi+\epsilon
 \sin(k\theta) \frac{\Pi}{R_o} \cosh \Pi].
\end{equation}
The calculated results are demonstrated in Figs. 3 and 4. The
parameters are fixed below (the orientation effect :
$\sin(k\theta)$ is fixed here). $r_o$ (the mean outer radius) is
selected as the same as the slip length $L_s^0=100$ nm. The
amplitude of wavy roughness is $\epsilon=0.04, 0.07, 0.1$,  the
Boltzmann constant ($k_B$) is $1.38 \times 10^{-23}$
Joule/$^{\circ}$K, and the Planck constant ($h$) is $6.626 \times
10^{-34}$ Joule $\cdot$ s. \newline \noindent In each panel, the
inner curve is the relevant boundary of the tube or the geometric
part of the presentation. The distance between the inner and
corresponding outer curves is the calculated physical shear rate :
$\dot{\xi}$.
%As the $y$-axis is a normalized value (with respect to $d\xi_0/dt$), thus
%it is dimensionless. The temperature effect is absent here.
We can observe once the temperature ($T$) changes a little  from
285 $^{\circ}$ to 295 $^{\circ}$, the enhancement of $\dot{\xi}$
becomes at least three orders of magnitude (for $\Pi=1$, the
activation energy : $3\times 10^{-23}$ Joules). Even at very low
temperature Fig. 4 gives very large strain rates which are
required to to obtain the
 necessary strain for plastic deformation.
 Thus, the constitutive relations is highly nonlinear
 at rather low temperature regime [9].
 It is worth pointing out that the Eyring model requires the
interaction between atoms in the direction perpendicular to the
shearing direction for the momentum transfer. This might explain
why our result is orientation dependent. The effect of
wavy-roughness will be significant once the forcing ($\Pi$) is
rather large (the maximum is of the order of magnitude of
$\epsilon [\Pi \tanh(\Pi)/R_o]$).
\newline
\noindent To be specific, we can illustrate the shear rate
($\dot{\xi}$) with respect to the temperature ($T$) once we
calculate $\dot{\xi}_0$ as the latter is temperature dependent
(but presumed roughness independent here) which could be traced
from equation (1). This is shown in Fig. 5. \newline Note that,
based on the rate-state Eyring model (of stress-biased thermal
activation), structural rearrangement is associated with a single
energy barrier (height) $E$ that is lowered or raised linearly by
a (shear) yield stress $\tau$. If the transition rate is
proportional to the plastic (shear) strain rate (with a constant
ratio : $C_0$; $\dot{\xi} =C_0 R_t$, $R_t$ is the transition rate
in the direction aided by stress), we have $\tau = E / V^*+( k_B T
/ V^*)
 \ln ( \dot{\xi} /C_0 \nu_0)$ or
\begin{equation}
%\hspace*{12mm} \mbox{or} \hspace*{12mm}
\tau = \frac{E}{V^*}+(\frac{ k_B T}{V^*})
 \ln (\frac{|\dot{\tau}| V^*}{\nu_0 k_B T}),
\end{equation}
where $V^* \equiv V_h$ is a constant called the activation volume,
$k_B$ is the Boltzmann constant, $T$ is the temperature, $\nu_0$
is an attempt frequency or transition rate [8,25], and
$\dot{\tau}$ is the stress rate. Normally, the value of $V^*$ is
associated with a typical volume required for a molecular shear
rearrangement. Thus, if there is a rather-small (plastic) flow (of
the glass) at low temperature environment then it could be related
to a barrier-overcoming or tunneling for shear-thinning matter
along the wavy-roughness (geometric valley and peak served as
atomic potential surfaces) in cylindrical micropores when the
wavy-roughness is present. Once the geometry-tuned potentials
(energy) overcome this barrier, then the tunneling (spontaneous
transport) inside wavy-rough cylindrical micropores occurs.
\newline
\noindent To examine the behavior of the shear rate at low
temperature regime, we calculate $\dot{\xi}_0$ and $\dot{\xi}$
($C_0 \nu_0=5\times 10^{10}$ s$^{-1}$) with respect to the
temperature $T$ and show the results in Fig. 5. For a selected
activation energy : $5\times 10^{-22}$ Joule or $\sim 10^{-3}$ eV
(a little bit smaller than the binding energy of $^3$He), we can
find a sharp decrease of shear rates around $T\sim 0.01
^{\circ}$K. Below this temperature, there might be nearly
frictionless transport of glassy matter. Note also that, according
to Cagle and Eyring [8], $V^*=3 V \delta \xi/2$ for certain
material during an activation event, where $V$ is the deformation
volume, $\delta\xi$ is the increment of shear strain.\newline
%------ZAMM
%Parameters remain the same as previous
%demonstrated Figures (e.g., $V^*=0.2$ nm$^3$, $r_o=100$ nm). The
%temperature ($T$) effect means the calculation is for fixed
%pressure gradient $-dp/dz \sim 6.0 \times 10^{10}$ (Pa/m) ($\tau_0
%\equiv \tau_0(T)$, $V^*=0.2$ nm$^3$) but the forcing ($\Pi$) as
%well as $\dot{\xi}_0$ is temperature dependent (through $\tau_0$).
%The $\Pi$ effect means the forcing is temperature independent (the
%same as those in Fig. 2), but $\dot{\xi}_0$ is temperature
%dependent. $\dot{\xi}_0$ is also shown for comparison as it is
%presumed roughness independent here. The activation energy
%($\Delta E$) is $10^{-22}$ Joule. In fact, all the results shown
%in this figure depend on $\dot{\xi}_0$ and are thus very sensitive
%to $\Delta E$.
%----------------------
If we select a (fixed) temperature, say, $T=0.1 ^{\circ}$K, then
from the expression of $\tau_0$, we can obtain the shear stress
$\tau$ corresponding to above forcings ($\Pi$) :
\begin{equation}
 \tau =\tau_0 \sinh^{-1} [\sinh(\Pi)+\epsilon
 \sin(k\theta) \frac{\Pi}{R_o} \cosh(\Pi)].
\end{equation}
There is no doubt that the orientation effect ($\theta$) is also
present for deformation kinetics of polymeric matter. For
illustration (shown in Fig. 6), we only consider the maximum case
: $|\sin(k\theta)|=1$. The trend of enhancement due to $\Pi$
(pressure-forcing) and $\epsilon$ (roughness) is similar to those
presented in Figs. 3 and 4. We remind the readers that, due to the
appearance of $\tau_0$, we fix the temperature to be the same and
the activation volume : $10^{-25}$ m$^3$.\newline
\noindent In fact, as shown in Fig. 6, the calculated (shear)
stress (which is directly linked to the resistance of the glassy
matter) also shows a sudden decrease around $T\sim 1.1 ^{\circ}$K
especially for the case of ($C_0 \nu_0=2\times 10^{9}$ s$^{-1}$).
Here, the activation volume ($V^*$ or $V_h$) is selected as $0.2$
nm$^3$ [25]. Thus, the nearly frictionless transport of the glassy
fluid at low temperature environment (relevant to the
supersolidity, cf. [26]) could be related to a barrier-overcoming
or tunneling for shear-thinning matter along the wavy-roughness
(geometric valley and peak served as atomic potential surfaces) in
cylindrical micropores when the wavy-roughness is present. Once
the geometry-tuned potentials (energy) overcome this barrier, then
the tunneling (almost frictionless transport) inside
wavy-rough cylindrical micropores occurs. \newline %The geometric and/or quantum
%confinemnet which is relevant to this barrier, however, is yet
%obscure.
\noindent We also noticed that, as described in [9], mechanical
loading lowers energy barriers, thus facilitating progress over
the barrier by random thermal fluctuations. The simplified Eyring
model approximates the loading dependence of the barrier height as
linear. This Eyring model, with this linear barrier height
dependence on load, has been used over a large fraction of the
last century to describe the response of a wide range of systems
and underlies modern approaches to sheared glasses. The linear
dependence will always correctly describe small changes in the
barrier height, since it is simply the first term in the Taylor
expansion of the barrier height as a function of load. It is thus
appropriate when the barrier height changes only slightly before
the system escapes the local energy minimum. This situation occurs
at higher temperatures; for example, Newtonian deformation
kinetics is obtained in the Eyring model in the limit where the
system experiences only small changes in the barrier height before
thermally escaping the energy minimum. As the temperature
decreases, larger changes in the barrier height occur before the
system escapes the energy minimum (giving rise to, for example,
non-Newtonian deformation kinetics). In this regime, the linear
dependence is not necessarily appropriate, and can lead to
inaccurate modelling. This explains why we should adopt the
hyperbolic sine law [9] to treat the glassy matter.
\newline
To be specific, our results are rather sensitive to the
temperature ($T$) and the activation energy. Fig. 7 shows
especially the temperature dependence of the forcing parameter
($\Pi$) if $dp/dz$ is prescribed (say, around $6\times 10^{10}$
Pa/m) and the activation volume is $0.2$ nm$^3$ ($r_o=100$ nm). We
can observe that once $T$ increases $\Pi$ decreases. $\dot{\xi}$
calculated using prescribed $\Pi$ and using directly $T$ also
differs. Finally, we present the calculated maximum velocity (unit
: m/s) with respect to the temperature in Fig. 8. Geometric
parameters : $r_o$ and the activation volume are the same as those
in Fig. 7 and the roughness amplitude $\epsilon=0.02, 0.05 R_o$.
We consider the effect of the activation energy : $9.0 \times
10^{-23}$,  $1\times 10^{-22}$ and $2\times 10^{-22}$ Joule.
Around $T\sim 0.35 ^{\circ}$K, the maximum velocity (of the glassy
matter) either keeps decreasing as the temperature increases for
larger activation energy  or instead increases as the temperature
increases for smaller activation energy! The results presented in
Fig. 8 might be related to the microscopic origin for physical
aging  or effects of thermal history [27] and indeterminate
solutions discussed in [28]. The latter observation might be
related to the argues raised in [29] for the annealing process of
solid helium at similar low temperature environment if we treat
the solid helium to be glassy at low temperature regime.
%To conclude in brief,
%we found the glass inside cylindrical micropores could flow without any applied forcing only when the micropore
%surface is not smooth or wavy-rough. The spontaneous flow rate is
%rather small (of the order of magnitude of the square of the small
%wavy-roughness amplitude) and is proportional to the (referenced)
%shear rate, the slip length and the difference between the outer
%mean radius and the inner mean radius times the cosine of the
%phase shift of the wavy-roughness as illustrated above.
\section{Conclusions}
To give a brief summary, we analytically obtain a class of
temperature as well as activation energy dependent fields of the
rate of deformation for glassy polymer in microscopically confined
wavy-rough domain at very low temperature regime. The effects of
wavy corrugation upon the confined deformation kinetics at
very-low temperature are clearly illustrated. It is found that
there exist almost frictionless plastic flow fields for the rate
of deformation of glassy polymer inside cylindrical micropores at
very low temperature once the micropore surface is wavy-rough and
the activation energy is prescribed. Once the temperature,
activation volume, and geometry are fixed, the increase of
activation energy instead reduces significantly the (maximum) rate
of deformation of the glassy matter. The critical rate of
deformation is proportional to the (referenced) shear rate, the
slip length, the orientation and the amplitude of the
wavy-roughness as illustrated above.  \newline
%{\it Acknowledgement.} The author is partially supported by the 2007-Heb-NU Starting Funds
%for Scientific Researcher.
%\subsection*{Acknowledgement.} The author is partially supported
%by the 2007-Heb-NU Starting Funds for Scientific Researcher. The
%author stayed at the Chern Shiing-Shen Institute of Mathematics,
%Nankai University around the beginning of 2008-Jan. Thus the
%author should thank their hospitality for the first stage
%Visiting-Scholar Program. \newline
%---------------------------------------------------
%
%As for the slip length $L_s^0$, we should take the {\it quantum
%slip effect}\cite{Q:SLip} into account because it is the
%quantum-mechanical scattering of Bogoliubov quasiparticles which
%is responsible for the loss of transverse momentum transfer to the
%container walls.
%--------------
%

%--------------

\newpage

\psfig{file=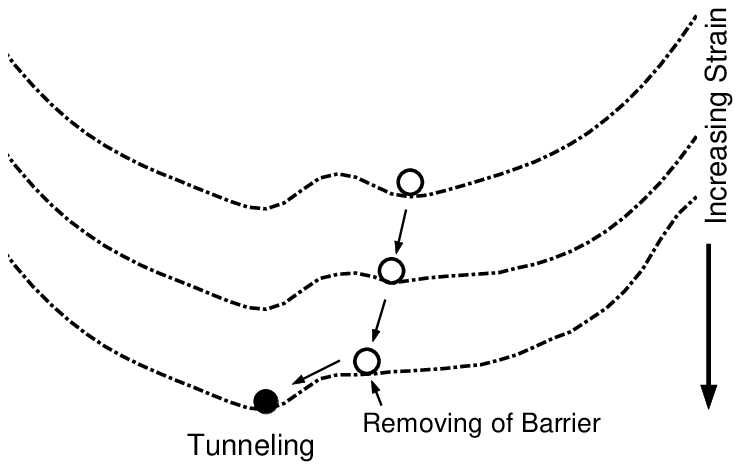,bbllx=-1.0cm,bblly=18cm,bburx=10cm,bbury=27cm,rheight=8cm,rwidth=8cm,clip=}

\begin{figure}[h]
\hspace*{10mm} Fig. 1 Increasing strain causes a local energy
minimum to flatten until it disappears \newline \hspace*{10mm}
(removing of energy barrier or quantum-like tunneling). The
structural contribution to \newline \hspace*{10mm} the shear
stress is shear thinning.
\end{figure}

\newpage
\psfig{file=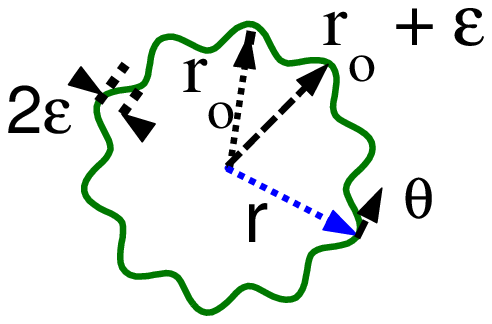,bbllx=1.0cm,bblly=13cm,bburx=18cm,bbury=17cm,rheight=5cm,rwidth=6cm,clip=}

\begin{figure}[h]
\hspace*{6mm} Fig. 2.  Schematic (plot) of a  micropore.
 $\epsilon$ is
the amplitude of small wavy-roughness.
\end{figure}

\newpage

\psfig{file=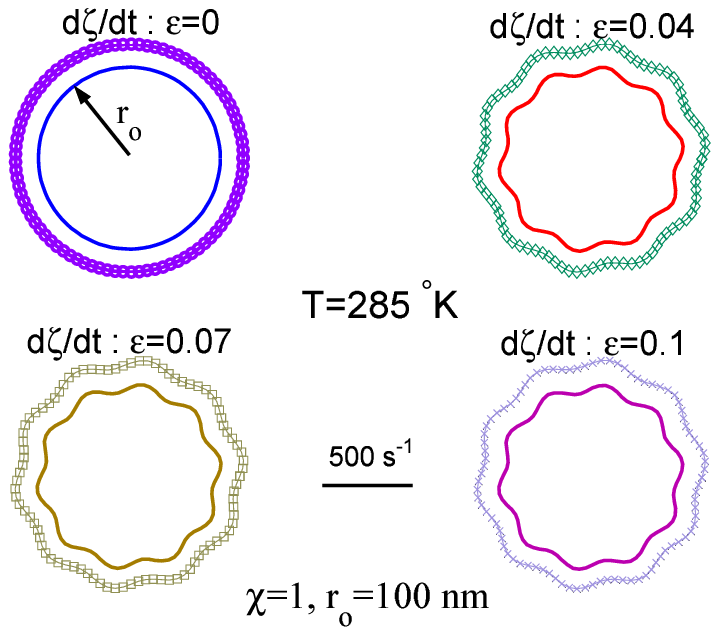,bbllx=-1.0cm,bblly=18cm,bburx=14cm,bbury=26cm,rheight=8cm,rwidth=10cm,clip=}

\begin{figure}[h]
%\vspace{22mm}
{\small Fig. 3. Comparison of the shear rate ($\dot{\zeta}$) of
polymeric matter in smooth microtubes and wavy-rough microtubes
for $k=10$, $\epsilon=0.0, 0.04, 0.07, 0.1$ with $r_o$=100 nm.
$\dot{\zeta_c}/\dot{\zeta}_0=10$ and $L^0_s=r_o$. $k$ is the wave
number and $\epsilon$ is the amplitude of the wavy-roughness. $T$
is the temperature. The solid-line length represents the scale of
$\dot{\zeta}=500 s^{-1}$.}

\end{figure}

\newpage

\psfig{file=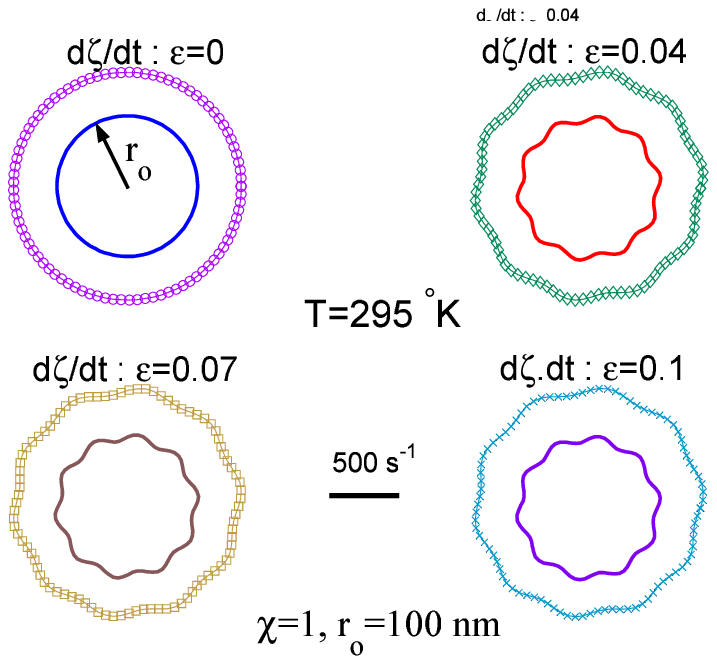,bbllx=-1.0cm,bblly=18cm,bburx=14cm,bbury=26cm,rheight=8cm,rwidth=10cm,clip=}
\begin{figure}[h]
%\vspace{22mm}
{\small Fig. 4. Comparison of the shear rate ($\dot{\zeta}$) of
polymeric matter in smooth microtubes and wavy-rough microtubes
for $k=10$, $\epsilon=0.0, 0.04, 0.07, 0.1$ with $r_o$=100 nm.
$\dot{\zeta_c}/\dot{\zeta}_0=10$ and $L^0_s=r_o$. $k$ is the wave
number and $\epsilon$ is the amplitude of the wavy-roughness. $T$
is the temperature. The solid-line length represents the scale of
$\dot{\zeta}=500 s^{-1}$.} As the temperature increases a little,
$\dot{\zeta}$ increases significantly.

\end{figure}

\newpage

\psfig{file=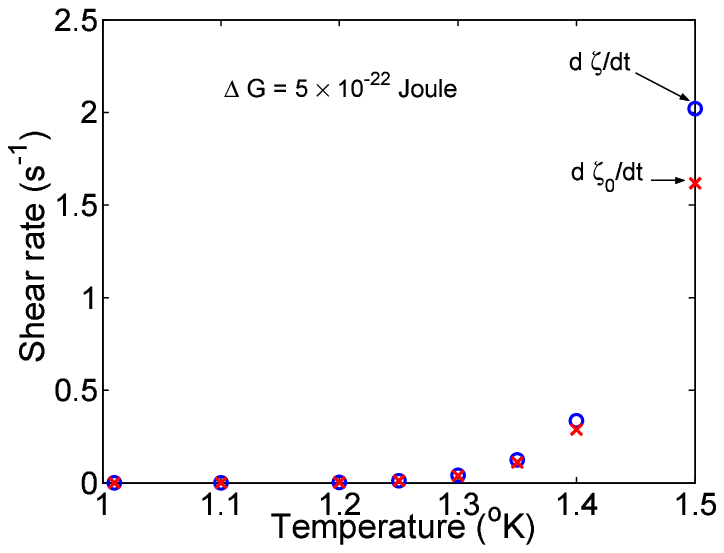,bbllx=-1.0cm,bblly=19cm,bburx=12cm,bbury=26.8cm,rheight=7cm,rwidth=9cm,clip=}
\begin{figure}[h]
\hspace*{10mm} Fig. 5 Comparison of calculated shear (strain)
rates using an activation energy $5\times 10^{-22}$ \newline
\hspace*{10mm} Joule or $\sim 10^{-3}$ eV. There is a sharp
decrease of shear rate around $T\sim 1.4 ^{\circ}$K. \newline
\hspace*{10mm} $C_0 \nu_0=5\times 10^{10} s^{-1}$ ($\nu_0 \sim
R_t$).
\end{figure}

\newpage

\psfig{file=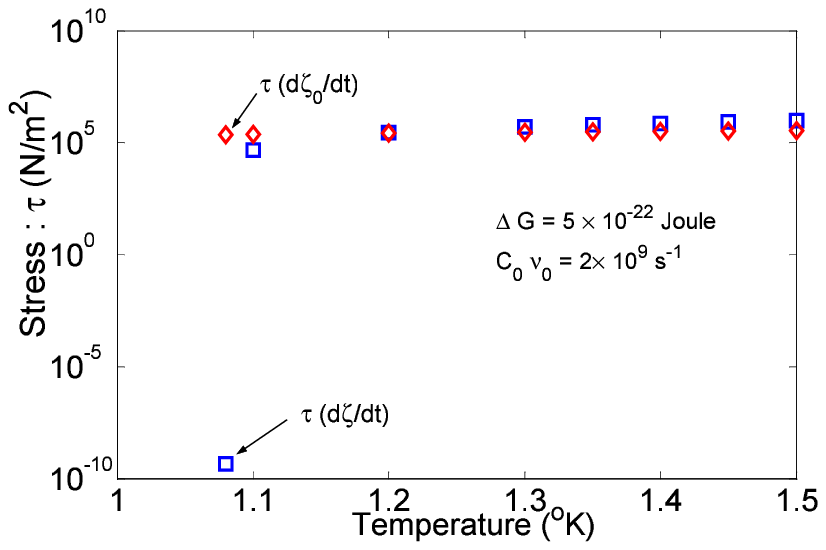,bbllx=-1.0cm,bblly=19cm,bburx=12cm,bbury=26.8cm,rheight=7cm,rwidth=8.5cm,clip=}

\begin{figure}[h]
\hspace*{10mm} Fig. 6  Comparison of calculated (shear) stresses
using an activation energy $5\times 10^{-22}$
\newline \hspace*{10mm} Joule or $\sim 10^{-3}$ eV. There is a
sharp decrease of shear stress around $T\sim 1.1 ^{\circ}$K for
\newline \hspace*{10mm}  $C_0 \nu_0=2\times 10^9$ s$^{-1}$. Below
$1.1  ^{\circ}$K, the transport of polymeric matter is nearly
frictionless. \newline \hspace*{10mm} $\nu_0$ is an attempt
frequency or transition rate$^{25}$.
\end{figure}

\newpage

\psfig{file=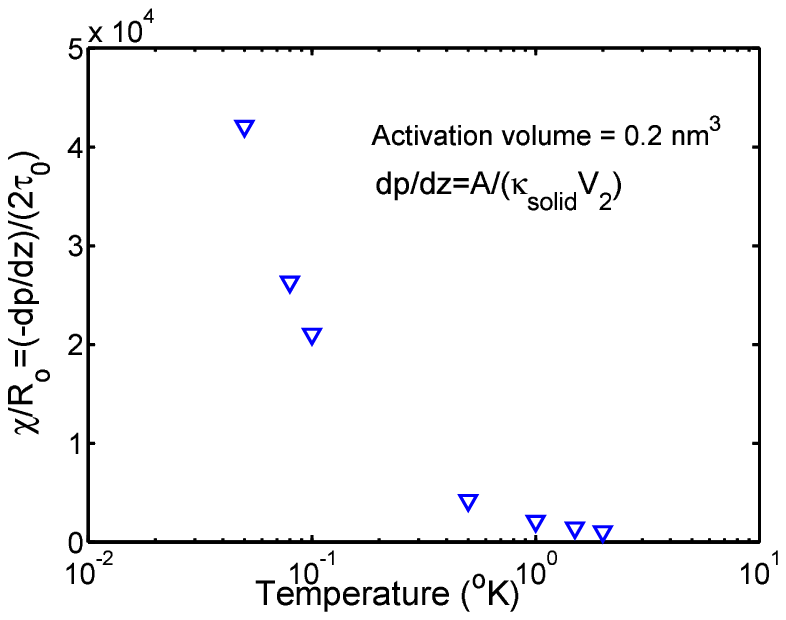,bbllx=-1.0cm,bblly=18.5cm,bburx=12cm,bbury=26.5cm,rheight=8cm,rwidth=9cm,clip=}

\begin{figure}[h]
\hspace*{10mm} Fig. 7 Calculated forcing parameters ($\chi/R_o$)
w.r.t. the temperature ($T$).\newline \hspace*{10mm}
$\kappa_{solid}$ is the compressibility, $A$ is the area, and
$V_2$ is the selected volume.
\newline \hspace*{10mm} Forcing ($\chi$) decreases as the
temperature ($T$) increases. %$\dot{\zeta}$ (mark : diamond)
%calculated \newline \hspace*{10mm} using prescribed $\chi$
% is different from that (mark : cross)
%using directly $T$ (temperature).
\end{figure}

\newpage

\psfig{file=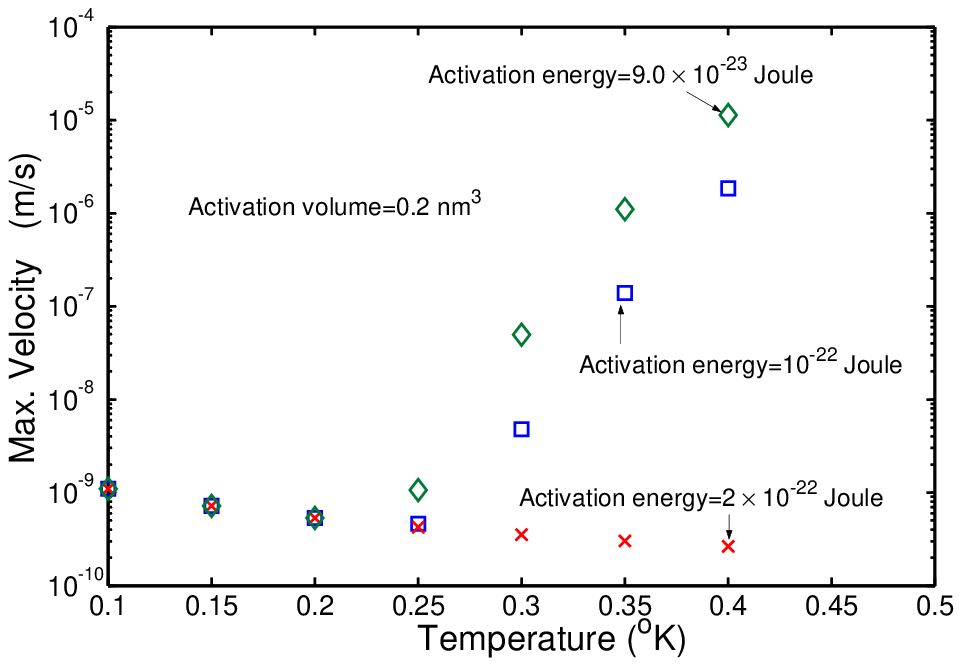,bbllx=-1.0cm,bblly=19.5cm,bburx=12cm,bbury=27cm,rheight=8.5cm,rwidth=9.5cm,clip=}

\begin{figure}[h]
\hspace*{10mm} Fig. 8  Comparison of calculated (maximum) velocity
(unit : m/s) using two activation \newline \hspace*{10mm} energies
$1.5 \times 10^{-22}$ and $2\times 10^{-22}$
 Joule. Around $T\sim 0.35 ^{\circ}$K, the monotonic trend of
  \newline \hspace*{10mm} velocity (or deformation rate)
   bifurcates as the temperature increases. \newline \hspace*{10mm}
  $r_o=100$ nm and $\epsilon=0.02, 0.05 R_o$.
\end{figure}

\begin{thebibliography}{99}
\bibitem{Plaist:2007}
T.A. Plaisted and S.  Nemat-Nasser,
%Quantitative evaluation of fracture, healing and re-healing of a
%reversibly cross-linked polymer
Acta Mater. {\bf 55}, 5684 (2007).  % -5696.
%{Arrat:2005}Arratia PE, Shinbrot T, Alvarez MM,  Muzzio FJ.
%Mixing of non-Newtonian \newline \hspace*{4mm} fluids in steadily forced
%systems.
%{Phys Rev Lett} 2005;{94}: 084501/1-4. %\newline
\bibitem{Boyce:1989}
R.C. Scogna and R.A.  Register, %Rate-dependence of yielding in ethylene每methacrylic acid
%copolymerscan
Polymer {\bf 49}, 992 (2008).  %-998.
%Boyce MC, Parks DM, Argon AS.
% Plastic flow in oriented glassy polymers.
%Int  J Plast 1989; 5: 593.  %-615. \newline
\bibitem{Jacobsen:2008}
A.J. Jacobsen, W.  Barvosa-Carter and S.  Nutt,
%Shear behavior of polymer micro-scale truss structures formed from
%self-propagating polymer waveguides
Acta Mater. {\bf 56}, 1209 (2008).  %-1218.
%{Daniel:2007}Danielsson M, Parks DM,  Boyce MC.
% Micromechanics, macromechanics and \newline \hspace*{4mm} constitutive modeling of the
%elasto-viscoplastic deformation of rubber-toughened glassy \newline \hspace*{4mm} polymers.
%J Mech Phys Solids 2007; 55: 533.  %-561.\newline
\bibitem{Haward:1997}
R.N. Haward and R.J. Young,
 The Physics of Glassy Polymers (Chapman and Hall, London, 1997).  %\newline
\bibitem{Oyen:2007}
S.A. Baeurle, A. Hotta and A.A.  Gusev,
%On the glassy state of multiphase and pure polymer materials
Polymer {\bf 47}, 6243 (2006).  %-6253.
% Oyen ML.
%Sensitivity of polymer nanoindentation creep measurements to
%experimental variables
%Acta Mater 2007; 55: 3633.  %-3639
%{Khan:2001}Khan A, Zhang Hy.
% Finite deformation of a polymer: experiemnts and modeling.
% Int J Plast 2001; 17: 1167.  %-1188. \newline
\bibitem{Jacobsen:2007}
J. Mohanraj, D.C. Barton, I.M. Ward, A. Dahoun, J.M. Hiver and C.
G'Sell,
%Plastic deformation and damage of polyoxymethylene in the large
%strain range at elevated temperatures
Polymer {\bf 47}, 5852 (2006).  %-5861.
%Jacobsen AJ, Barvosa-Carter W,  Nutt S.
%Compression behavior of micro-scale truss structures formed from
%self-propagating polymer waveguides
%Acta Mater 2007; 55: 6724.  %-6733
%{Kobel:2005}Kobelev V, Schweizer KS.
% Strain softening, yielding, and shear thinning in
%glassy \newline \hspace*{4mm} colloidal suspensions.
%Phys Rev E 2005;71: 021401/1-4. %\newline
\bibitem{Paint:1997}
P.C. Painter and M.M. Coleman,  Fundamentals of Polymer Science:
An Introductory  Text  (Technomic Publishing Co., Inc. 2nd. ed.,
New York, 1997).
\bibitem{Cagle:1951}
F.W. Cagle Jr. and H. Eyring,
%An application of the absolute reaction rate theory to
%some problems in annealing.
{J. Appl. Phys.} {\bf 22}, 771 (1951). %-775.
\bibitem{Eyring:1936}
H. Eyring,
% Viscosity, plasticity, and diffusion as examples of absolute reaction rates.
{J. Chem. Phys.} {\bf 4}, 283 (1936).  %-291. %\newline
A. Tobolsky and H. Eyring,
%Mechanical Properties of Polymeric Materials.
J. Chem. Phys. {\bf 11}, 125 (1943). %每134.
\bibitem{Falk:2004}
C.P. Buckley and D.C. Jones, %Glass-rubber constitutive model for amorphous polymers near the
%glass transition
Polymer {\bf  36}, 3301 (1995).  %-3312.
%Falk ML, Langer JS, Pechenik L.
% Thermal effects in the shear-transformation-\newline \hspace*{4mm} zone theory of
%amorphous plasticity: Comparisons to metallic glass data.
%Phys Rev E 2004; 70: 011507/1-11.  %\newline
\bibitem{Gueguen:2008}
O. Gueguen, J.  Richeton, S.  Ahzi and A.  Makradi,
%Micromechanically based formulation of the cooperative model for
%the yield behavior of semi-crystalline polymers
Acta Mater. {\bf  56}, 1650 (2008). %-1655.
%{Lion:1997}Lion A.
% On the large deformation behaviour of reinforced rubber at different \newline \hspace*{4mm} temperatures.
%J Mech Phys Solids 1997;45: 1805.  %-1834.\newline
\bibitem{Liu:2001}
A.J. Liu and S.R. Nagel, Jamming and Rheology (Taylor \&
Francis,   London, 2001).  %\newline
\bibitem{Mayr:2006}
K.C. Valanis and S.T.J.  Peng, %Deformation kinetics of ageing materials
Polymer {\bf 24}, 1551 (1983).  %-1557.
%Mayr SG.
% Activation energy of shear transformation zones: a key for understanding
%\newline \hspace*{4mm} rheology of glasses and liquids.
%Phys Rev Lett 2006; 97: 195501/1-4.  %\newline
\bibitem{Haxton:2007}
T.K. Haxton and A.J.  Liu,
% Activated dynamics and effective temperature in a steady \newline \hspace*{4mm} state sheared glass.
Phys. Rev. Lett. {\bf  99}, 195701 (2007).  %\newline
\bibitem{Kacz:2003}
T. Pietsch, N.  Gindy, A. Fahmi, %Preparation and control of functional nano-objects: Spheres, rods
%and rings based on hybrid materials
Polymer {\bf 49}, 914 (2008).  %-921.
%Kaczmarek J.
% A nanoscale model of crystal plasticity.
%Int J Plast 2003; 19: 1585.  %-1603. \newline
\bibitem{Li:2008}
J.-L. Yang, Z. Zhang, A.K. Schlarb and K. Friedrich, %On the characterization of tensile creep resistance of polyamide
%66 nanocomposites. Part II: Modeling and prediction of long-term
%performance
Polymer {\bf 47}, 6745 (2006).  %-6758.
%Li JY, Mueller J, Ho\"{o}ppel HW,
% G\"{o}ken M,  Blum W.
%  Deformation kinetics of \newline \hspace*{4mm} nanocrystalline nickel.
%Acta Mater  2007;55: 5708.  %-5717. \newline
\bibitem{Bell:1985}
J.F. Bell,
% Contemporary perspectives in finite strain plasticity.
Int. J. Plast. {\bf 1}, 3 (1985). %-27.
J.F. Bell,
% Plane stress, plane strain, and pure shear at large finite strain.
Int. J. Plast.  {\bf 4}, 127 (1988). %-148.
\bibitem{Chu:2007}
Z. K.-H. Chu,
% Rapid transport of glassy supersolid helium in wavy-rough nanpores. \newline \hspace*{4mm}
Arxiv:0707.2828. %\newline
\bibitem{Tijum:2007}
R. van Tijum, W.P. Vellinga and J.Th.M. De Hosson,
%Surface roughening of metal每polymer systems during plastic
%deformation
Acta Mater.{\bf 55}, 2757 (2007).  %-2764.
%{Hiratani:2003}
%Hiratani M,  Zbib HM, Khaleel MA.
% Modeling of thermally activated dislocation \newline \hspace*{4mm} glide and plastic flow through
%local obstacles.
%Int J Plast 2003; 19: 1271.  %-1296.\newline
\bibitem{Trias:2006}
D. Trias, J.  Costa, A.  Turon and J.E.  Hurtado,
%Determination of the critical size of a statistical representative
%volume element (SRVE) for carbon reinforced polymers
Acta Mater. {\bf  54},  3471 (2006).  %-3484.
%{Balke:1994}
%Balke H, Estrin Y.
%Micromechanical modelling of shear banding in single crystals.
%Int J Plast 1994; 10: 133.  %-147. \newline
\bibitem{Deseri:2000}
F. Perrin-Sarazin, M.-T. Ton-That, M.N. Bureau and J.  Denault,
%Micro- and nano-structure in polypropylene/clay nanocomposites
Polymer {\bf  46}, 11624 (2005).  %-11634.
%Deseri L, Owen DR.
% Active slip-band separation and the energetics of slip in \newline \hspace*{4mm} single crystals.
%Int J Plast 2000; 16: 1411.  %-1418. \newline
%Bardenhagen, S.G., Stout, M.G., Gray, G.T., 1997.
%Three-dimensional, finite deformation, \newline \hspace*{4mm} viscoplastic constitutive models for
%polymeric materials. Mech. Mater. 25, 235-253.\newline
%Borg, U., 2007. Strain gradient crystal plasticity effects on flow
%localization. Int. J. Plast. \newline \hspace*{4mm} 23, 1400-1416.
%\newline
%Br\"{u}nig, M. and  Obrecht, H., 1998. Finite elastic-plastic
%deformation behaviour of \newline \hspace*{4mm} crystalline solids
%based on a non-associated macroscopic flow rule. Int. J. Plast.
%14, \newline \hspace*{4mm} 1189-1208.\newline
\bibitem{Johnson:1997}
K.L. Johnson,
% Adhesion and friction between a smooth elastic spherical asperity \newline \hspace*{4mm} and a plane surface.
Proc. Royal Soc.  London A {\bf 453}, 163 (1997). %-179.
\bibitem{Chu:1996}
W. K.-H. Chu,
%Stokes slip flow between corrugated walls.
{Z. Angew. Math. Phys.} {\bf 47}, 591 (1996).  %-599.
\bibitem{Taylor:1938}
G.I. Taylor,  %Plastic strains in metals.
J. Inst. Met. {\bf 62}, 307 (1938). %Proc. Roy. Soc 50, 307.\newline
\bibitem{Capaldi:2004}
F.M. Capaldi, M.C. Boyce and G.C. Rutledge,
%Molecular response of a glassy polymer \newline \hspace*{4mm} to active deformation.
Polymer {\bf 45}, 1391 (2004).  %-1399.
\bibitem{Pomeau:1994}
Y. Pomeau and S. Rica,
% Dynamics of a model of a supersolid.
Phys. Rev. Lett. {\bf 72}, 2426 (1994).  %-2430.
\bibitem{Aging:Nano}
A. Wypych, E.  Duval, G. Boiteux, J.  Ulanski, L. David and A.
Mermet,
%Effect of physical aging on nano- and macroscopic properties of
%poly(methyl methacrylate) glass
Polymer {\bf 46}, 12523 (2005). %-12531.
\bibitem{Stoughton:2008}
T.B. Stoughton and J.W. Yoon,
% On the existence of indeterminate solutions to the \newline \hspace*{4mm} equations of motion under
%non-associated flow.
Int. J. Plast. {\bf 24}, 583 (2008).  %-613. %\newline
%  Anderson, P.W. Bose fluids above Tc: Incompressible vortex fluids
%  and "Supersolidity". Nature Phys. 3, 160-162  (2007); cond-mat/07051174.
%  Andreev, A.F. Supersolidity of glasses, cond-mat/07050571.
%\bibitem{Day:Melt}
%Flow of Solid 4He Near Melting
%J. Day and J. Beamish,
%J. Low Temp. Phys.  {\bf 148}, 683 (2007).  %每687
%\bibitem{Visual:S-He}
%Visual Observations of Disordered Solid Helium
%N.C. Ford, Jr.,  R.B. Hallock and K.H. Langley, J. Low Temp. Phys.
%{\bf 148}, 653 (2007).  %每657
%\bibitem{Eyring:JChemP}
%H. Eyring, J. Chem. Phys. {\bf 4}, 283 (1936).
%\bibitem{E:Barrier}
%F.W. Cagle, Jr. and H. Eyring, J. Appl. Phys.  {\bf 22},  771  (1951).  %-775
%C.E. Maloney and D.J. Lacks, Phys. Rev. E {\bf 73}, 061106 (2006).
%\bibitem{S:Thin-Glass} J. Rottler and M.O. Robbins, Phys. Rev. E
%{\bf 68}, 011507 (2003).
%\bibitem{Slip:SF}
%G.J. Hyland and G. Rowlands,
%On the microscopic derivation of the two-fluid thermohydrodynamic
%equations for4He II
%J. Low Temp. Phys. {\bf 7}, 271 (1972).  %每289
%\bibitem{Q:SLip}
%6.  Einzel, D. \&  Parpia, J.M. Slip in quantum fluids. {\it J.
%Low Temp. Phys.} {\bf 109}, 1-105 (1997). (cf. the quantum slip
%effect at page 40 therein)  \newline
%%-105 (1997).
%\bibitem{F:SLip}
%Boundary effects in fluid flow. Application to quantum liquids
%H.H. Jensen, H. Smith, P. W\"{o}lfle,  K. Nagai and T.M. Bisgaard,
%J. Low Temp. Phys. {\bf 41}, 473 (1980).  %-519
%\bibitem{B:Perturb} W. K.-H.
%Chu, {ZAMP} {\bf 47}, 591 (1996).
%Onuki, A., 2003. Plastic flow in two-dimensional solids. {Phys.
%Rev. E} {68}, 061502/1-10.\newline
%Ostoja-Starzewski, M., 2005.
%Scale effects in plasticity of random media: status and
%\newline \hspace*{4mm} challenges. Int. J. Plast. 21, 1119-1160.  \newline
%\newline
\bibitem{Penzev:2007}
A. Penzev, Y.  Yasuta and M.  Kubota,
%Annealing effect for supersolid fraction in $^4$He. \newline \hspace*{4mm}
J. Low Temp. Phys. {\bf 148}, 677  (2007). %-681.

\end{thebibliography}
\end{document}